\documentclass[lettersize,journal]{IEEEtran}
\usepackage{amsmath,amsfonts}
\usepackage{algorithmic}
\usepackage{algorithm}
\usepackage{array}
\usepackage[caption=false,font=normalsize,labelfont=sf,textfont=sf]{subfig}
\usepackage{textcomp}
\usepackage{stfloats}
\usepackage{url}
\usepackage{verbatim}
\usepackage{graphicx}
\usepackage{cite}
\usepackage{xcolor}
\begin{document}

\title{Two-layer consensus based on master-slave \\ consortium chain data sharing for Internet of Vehicles}

\author{Feng Zhao,~\IEEEmembership{Member,~IEEE, }Benchang Yang, Chunhai Li,~\IEEEmembership{Member,~IEEE, }Chuan Zhang,~\IEEEmembership{Member,~IEEE, }\\ Liehuang Zhu,~\IEEEmembership{Senior Member,~IEEE, }and Guoling Liang
\thanks{This work was supported in part by the Guangxi Natural Science Foundation under Grant 2023GXNSFAA026294; in part by the National Natural Science Foundation of China under Grant 62362013, Grant 62232002 and Grant 62202051; and in part by the Yunnan Provincial Major Science and Technology Special Plan Projects under Grant 202302AD080003. \textit{(Corresponding author: Chunhai Li.)}}
\thanks{Feng Zhao, Benchang Yang and Chunhai Li are with the Guangxi Engineering Research Center of Industrial Internet Security and Blockchain, Guilin University of Electronic Technology, Guilin 541004, China (e-mail: zhaofeng@guet.edu.cn; yangbc0124@163.com; chunhaili@guet.edu.cn).}
\thanks{Chuan Zhang and Liehuang Zhu are with the School of Cyberspace Science and Technology, Beijing Institute of Technology, Beijing 100081, China (e-mail: chuanz@bit.edu.cn; liehuangz@bit.edu.cn).}
\thanks{Guoling Liang is with the School of Information and Communication, Guilin University of Electronic Technology, Guilin 541004, China (e-mail: guolliang@mails.guet.edu.cn).}
}

\markboth{Journal of \LaTeX\ Class Files,~Vol.~14, No.~8, August~2021}%
{Shell \MakeLowercase{\textit{et al.}}: A Sample Article Using IEEEtran.cls for IEEE Journals}

\IEEEpubid{\begin{minipage}{\textwidth}\ \centering
		Copyright \copyright 2024 IEEE. Personal use of this material is permitted. \\
		However, permission to use this material for any other purposes must be obtained 
		from the IEEE by sending a request to pubs-permissions@ieee.org.
\end{minipage}}

\maketitle

\begin{abstract}
Due to insufficient scalability, the existing consortium chain cannot meet the requirements of low latency, high throughput, and high security when applied to Internet of Vehicles (IoV) data sharing. Therefore, we propose a two-layer consensus algorithm based on the master-slave consortium chain - Weighted Raft and Byzantine Fault Tolerance (WRBFT). The intra-group consensus of the WRBFT algorithm adopts weighted Raft, and the best node is selected as the master node to lead the intra-group consensus by comprehensively evaluating the signal-to-noise ratio (SNR), data processing capacity and storage capacity of the nodes. The inter-group consensus adopts practical Byzantine fault tolerance (PBFT) based on BLS aggregate signature with nonlinear coefficients to ensure that the inter-group consensus can tolerate 1/3 of Byzantine nodes. At the same time, the verifiable random function (VRF) is used to select the master node of the inter-group consensus to ensure the randomness of the master node. A large number of experimental results show that the proposed WRBFT algorithm reduces delay, and improves throughput and system security.
\end{abstract}


\section{Introduction}
\IEEEPARstart{T}{he} Internet of Vehicles (IoV) has become a new technology to support intelligent driving and improve traffic services\cite{ref1, ref2}. IoV data sharing can promote the intelligence, efficiency, and automation of the IoV system, and help vehicle managers better grasp vehicle operation and road information\cite{ref3}. In the data-sharing process of the IoV, roadside units (RSU) play an important role in data interaction with vehicles or cloud servers. Due to the lack of perfect security measures, RSU is easy to become the target of malicious attacks, resulting in data leakage or malicious tampering\cite{ref4, ref5}. Therefore, improving the security of RSU in data sharing has become the focus of attention\cite{ref6}. In recent years, with its characteristics of decentralization, auditability, traceability, and anonymity, the blockchain has provided a data security solution for the research of the IoV\cite{ref7, ref8, ref9}. By applying blockchain technology, a safe and efficient intelligent transportation system (ITS)\cite{ref10} can be built to improve the security of data sharing.

The consortium chain has received widespread attention due to its low transaction cost, fast transaction execution speed, and excellent privacy protection characteristics, which enables it to effectively meet the needs of RSU in data sharing. However, the current consortium chain system still has problems, and it is difficult to meet the low latency, high throughput, and excellent security requirements in IoV data-sharing scenarios. The consensus algorithm is a key method to ensure data consistency and improve data sharing efficiency, so many scholars have improved the consensus algorithm in the consortium chain. The PBFT\cite{ref11} consensus algorithm has Byzantine fault tolerance, but the high communication complexity has become the main factor that curbs its performance. These papers\cite{ref12, ref13} divide the nodes into multiple layers, and different layers independently perform PBFT consensus work, which effectively improves the efficiency of blockchain consensus. Zhang et al.\cite{ref14} proposed a data-sharing and storage system architecture based on consortium chains. The PBFT algorithm is used to increase the speed of data processing and the incentive mechanism is used to encourage vehicles to share data to ensure the stable operation of the IoV system. Lao et al.\cite{ref15} proposed a location-based and scalable PBFT algorithm, which mainly achieves consensus through the geographic location of fixed-location devices. Xu et al.\cite{ref16} proposed the SG-PBFT algorithm for the IoV, and adopted a fractional grouping mechanism to achieve higher consensus efficiency. At the same time, there are many excellent works\cite{ref17, ref18, ref19} that use blockchain technology to solve the problem of IoV data sharing. Although the above-mentioned PBFT consensus scheme improves the problem of low PBFT consensus efficiency to a certain extent, due to the lack of consideration of the openness of the IoV network, when the number of consensus nodes increases, the information density of the blockchain system increases exponentially.

\IEEEpubidadjcol

The Raft\cite{ref20} consensus algorithm is another consensus mechanism, which has the characteristics of low communication complexity and high throughput and is regarded as a solution to data sharing. Xu et al.\cite{ref21} proposed a weighted Raft consensus algorithm for the Internet of Things, which can reduce the system forwarding delay by up to 24\%. Based on the Raft consensus mechanism, Xu et al.\cite{ref22} studied the security performance of wireless blockchain networks under malicious interference and provided theoretical guidance for the actual deployment of wireless blockchain networks. Hou et al.\cite{ref23} proposed a smart transaction migration scheme based on the Raft consensus mechanism, which effectively reduces the data processing delay by migrating the transactions from the busy area to the idle area. Xue et al.\cite{ref24} proposed a decentralized fraud-proof roaming authentication framework based on blockchain and leverage smart contracts to implement a roaming authentication protocol, including user/AP registration, authentication, and revocation. In addition, there are some works\cite{ref25, ref26, ref27, ref28, ref29} that use different consensus mechanisms to solve the IoV data-sharing problem. Although the Raft consensus has the characteristics of low latency and high throughput, it lacks Byzantine fault tolerance. When improving the security of PBFT and Raft consensus, some works \cite{ref30, ref31, ref32, ref33} adopted VRF\cite{ref34} to select the master node, which improved the security of the algorithm. At the same time, there are also some works \cite{ref35, ref36, ref37} that use the ring signature cryptographic scheme to ensure the security of the blockchain. Most of the existing consortium chains use centralized or non-parallel consensus algorithms, which seriously affects the efficiency of data sharing. 

To enhance the efficiency and scalability of the data-sharing system, we propose a two-layer consensus mechanism for IoV data sharing based on master-slave consortium chains. In distributed systems, traditional consensus algorithms may face challenges such as a large number of nodes and high network communication latency. By utilizing a two-layer consensus, we can divide the network into different layers, each employing a different consensus algorithm. This approach allows for better adaptation to variations in communication latency between nodes and enhances the overall system performance. Additionally, the two-layer consensus can provide higher security as different consensus algorithms can complement each other's weaknesses, thereby reducing the risk of attacks on the system. Therefore, adopting a two-layer consensus for data sharing can effectively address the challenges faced by traditional consensus algorithms, improving the reliability and stability of the system.

The main contributions are presented as follows.

\begin{enumerate}


\item To the best of our knowledge, WRBFT is the first two-layer consensus algorithm designed using greedy thinking and is used for the data-sharing process in the IoV scenario.

\item We use a novel weighted Raft algorithm in the intra-group consensus of WRBFT, which comprehensively evaluates the node's SNR, data processing capacity, and storage capacity to select the master node in the group.

\item We developed a PBFT algorithm based on BLS aggregated signatures with non-linear coefficients in the inter-group consensus of WRBFT and used VRF to guarantee the randomness of the master node selection between groups.

\item Finally, a series of simulation experiments are carried out to verify the feasibility and effectiveness of the WRBFT algorithm in data sharing in the IoV.

\end{enumerate}

The rest of this paper is organized as follows. In the second section, the IoV data-sharing model based on the master-slave consortium chain is introduced, the WRBFT algorithm is developed in the third and fourth sections, and the performance evaluation of the algorithm is provided in the fifth section.

\section{System Design}
\noindent To ensure the efficiency and security of the IoV data sharing, we combined the actual needs of the IoV to design the data sharing model of the IoV based on the master-slave consortium chain. The consortium chain is composed of multiple pre-selected nodes with access rules blockchain\cite{ref38, ref39}. We choose the RSUs as the consensus node and deploy the consortium chain at the RSU, which is responsible for the consensus work of the entire blockchain.

\begin{figure}[!t]
\centering
\includegraphics[width=3.5in]{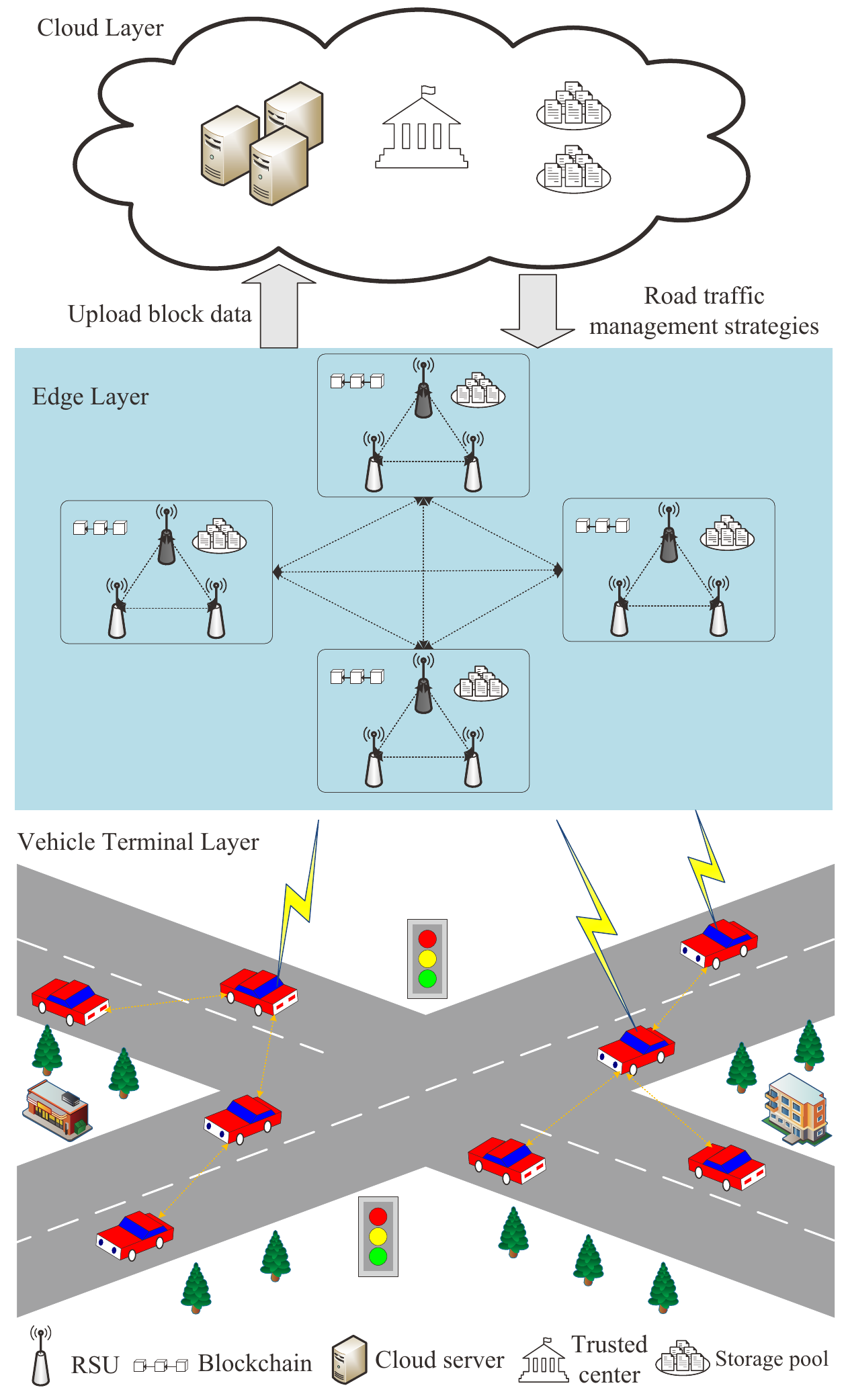}
\caption{Data sharing model of IoV based on master-slave consortium chain.}
\label{fig_1}
\end{figure}

\subsection{System Model}
\noindent The IoV data-sharing model we proposed based on the master-slave consortium chain is shown in Fig. 1, which is mainly composed of the cloud layer, the edge layer, and the vehicle terminal layer. The IoV data-sharing model combines the characteristics of blockchain decentralization, auditability, and traceability to effectively ensure the secure sharing and storage of vehicle data. We deploy the WRBFT algorithm at the edge layer and use the K-means clustering algorithm to group consensus nodes according to their geographic location. Each group runs the consensus work in parallel, and the master node in the intra-group consensus participates in inter-group consensus. Since the inter-group consensus adopts the PBFT consensus based on the BLS aggregate signature with nonlinear coefficients, the Byzantine fault tolerance of the IoV blockchain is guaranteed.

The basic functions of each layer in the IoV data sharing model are described as follows,

\begin{itemize}
\item Vehicle terminal layer: The main components are various types of vehicles, and their basic functions include obtaining their own basic information or road condition information through sensing units, cameras, and radio frequency identification units. The vehicle registers basic information through the trusted center (TA) and joins the IoV blockchain system as a data provider or data requester.
\item Edge layer: The main component is the RSUs. In addition to performing its basic functions, it also performs data interaction with vehicle terminals and RSUs within its communication range to obtain the latest vehicle information. As a blockchain consensus node, the RSUs are responsible for data collection, block generation, broadcast blocks, verification blocks, and block chaining during the operation of the WRBFT algorithm.
\item Cloud layer: The main component is a high-performance cloud server, which finally processes the received vehicle data through massive data computing and storage capabilities. The trained global model is sent to the RSU and vehicles, to better predict and optimize the vehicle driving route and road traffic management strategy in the next stage.
\end{itemize}

\subsection{Data Sharing Process}
\noindent When the vehicle passes the identity authentication of TA and joins the IoV blockchain system, the vehicle will have two identities of the data provider and the data requester. If the vehicle joins the blockchain system as a data owner, it will collect its own data or road traffic data regularly, and use the private key to pair the original data (speed, fuel consumption, GPS, traffic conditions, parking lot occupancy, etc.)\cite{ref40}) to sign, encrypt the public key and signature and upload it to the nearby or low-load RSUs. The RSUs decrypts the data after obtaining the authorization of the vehicle and uses the public key in the data to verify the signature. If the verification is passed, the original data is placed in the storage pool, and the storage address is returned. If the vehicle joins the blockchain system as a data requester, it will initiate a data request to the nearest or less loaded RSU within the communication range. The RSU returns the ID of the data owner. After obtaining the authorization information from the data owner, the vehicle requesting the data can legally access the data.

When the RSU receives a rated amount of data or exceeds the timeout period, it runs the WRBFT algorithm. First, the RSUs are evenly grouped according to their geographical location, and the RSUs in the intra-group consensus run the weighted Raft consensus algorithm to ensure the high efficiency of the blockchain system and select the master node in the intra-group consensus to enter the inter-group consensus. With the help of the idea of a greedy algorithm, the consensus nodes in the intra-group consensus are always the local optimum of each group and finally reach the global optimum after combination. It can be seen that the WRBFT algorithm can ensure that the entire IoV blockchain system has a lower delay, higher throughput, and security. The specific vehicle data-sharing process is shown in Fig. 2.

\begin{figure}[!t]
\centering
\includegraphics[width=3.5in]{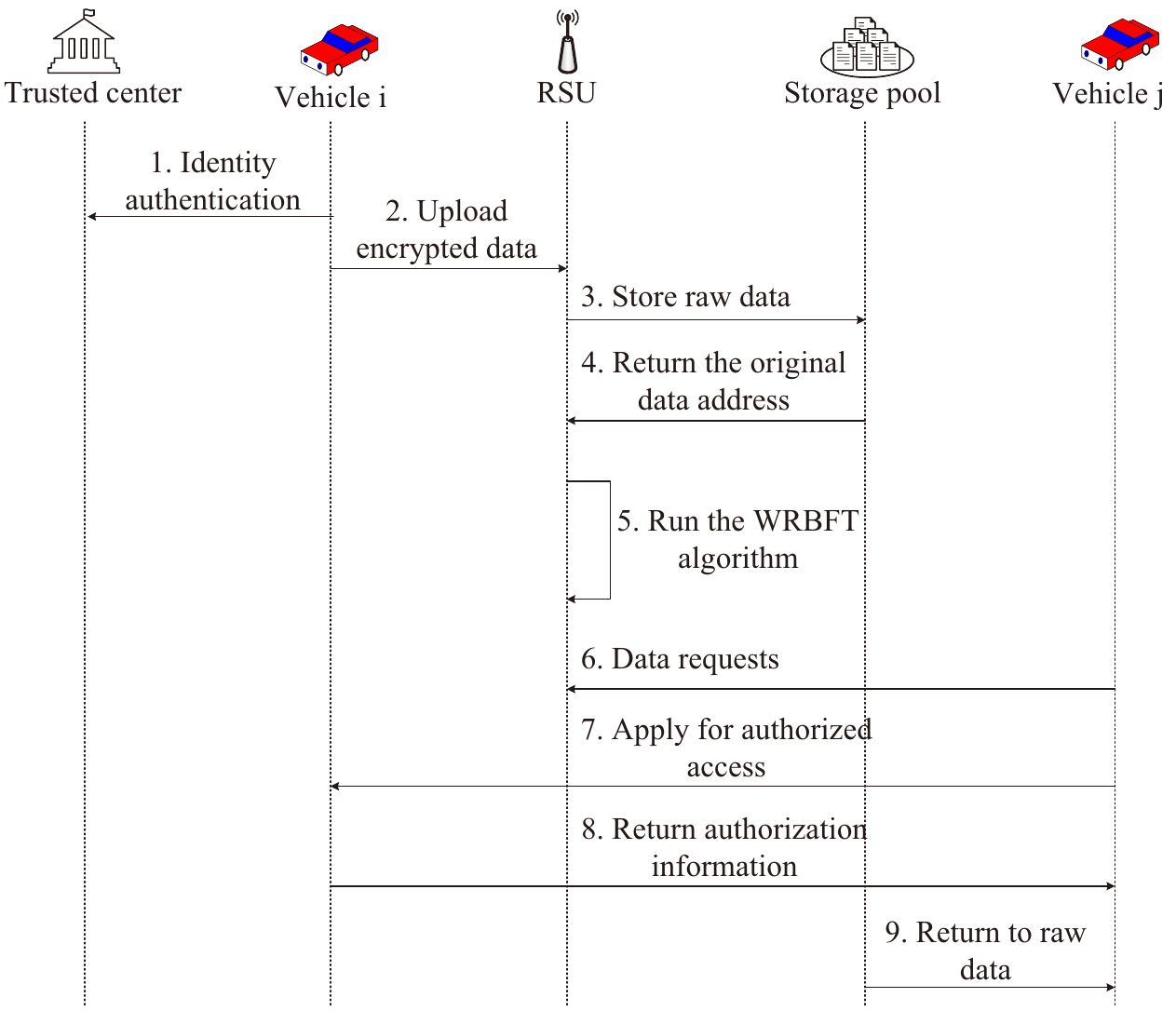}
\caption{IoV blockchain data sharing process.}
\label{fig_1}
\end{figure}

\section{Intra-group consensus of WRBFT algorithm}
\subsection{Consensus Node Grouping}
\noindent To make the WRBFT algorithm have lower latency, higher throughput, and security in the IoV blockchain system, we first use the K-means clustering algorithm to evenly group the consensus nodes, and each group independently runs the consensus within the group, and use the Euclidean distance between consensus nodes as the grouping basis, expressed as follows,
\begin{equation}
\label{1}
d\left ( i,j \right ) =\sqrt{\left ( x_{i} - x_{j} \right )^{2} + \left ( y_{i} - y_{j} \right )^{2}} 
\end{equation}
where the position coordinate of node $i$ is $\left ( x_{i},y_{i}\right ) $, and the position coordinate of node $j$ is $\left ( x_{j},y_{j}\right ) $.

Since the K-means clustering algorithm does not specify the number of nodes in each group, there will be a large difference in the consensus delay and throughput of each group. To balance the consensus delay and throughput of each group and for the sake of simplicity, we use a uniform grouping method to optimize the grouping strategy.

\begin{table}[!t]
\caption{Summary of Notations\label{tab:table1}}
\centering
\begin{tabular}{c|c}
\hline \textbf{Symbol} & \textbf{Description} \\
\hline $i$ & replica node id \\
$N$ & network size \\
$K$ & number of WRBFT consensus groups \\
$f$ & number of Down or Byzantine nodes \\
$v$ & view number \\
$h$ & block hash value \\
$data$ & vehicle data \\
$isVote$ & the sign of leader election \\
$\sigma _{i} $ & signature of message by replica node $i$ \\
$p$ & a large prime number \\
$G_{1} ,G_{2} $ & additive cyclic group of order $p$ \\
$g_{1} ,g_{2} $ & generator of $ G_{1} ,G_{2} $ \\
$e$ & bilinear mapping function \\
$G_{T} $ & additive cyclic groups of the same order as $G_{1} $ and $G_{2} $ \\
$H$ & secure hash function, $H:\left \{ 0,1 \right \} \to G_{1} $ \\
$m$ & consensus message\\
$\left ( pk_{i},sk_{i}   \right ) $ & key pair of replica node $i$\\
$\alpha _{i} $ & coefficient of replica node $i$ \\
$\hat{\sigma } $ & aggregate signature\\
$k$ & security parameters\\
$\xi ,\pi $ & random numbers and proof of random numbers\\
\hline
\end{tabular}
\end{table}

\subsection{Intra-group consensus}
\noindent The Raft consensus has the high efficiency of the Paxos \cite{ref41, ref42} algorithm, and has the advantages of low communication overhead, low delay, and high throughput, so it has become the first choice to coordinate intra-group consensus. Our in-depth research on the Raft consensus found that since the election of the leader node depends on the heartbeat timeout period, the node with a short heartbeat timeout period broadcasts election information faster, and it is easier to collect more than half of the election information and reply to become a leader. The randomness of the heartbeat time is not suitable for us to choose a node with better performance to become the leader. We propose to use a parameter related to node performance to affect the timeout time to improve the Raft consensus, that is, the weighted Raft consensus.

The wireless communication environment of each RSU is determined by the average SNR between the RSU and other RSUs when messages are forwarded. The total number of RSUs is expressed as $N$, and the calculation method of the wireless communication environment of RSU $i$ is as follows,
\begin{equation}
\label{3}
\overline{SNR_{i} } = \frac{1}{N - 1} \sum_{j=1,j\ne i}^{N} SNR_{i,j} 
\end{equation}

To comprehensively evaluate the RSU in the intra-group consensus, we will comprehensively evaluate the data processing capacity of the RSU, average SNR, and storage capacity to design a weight evaluation formula for the RSU,
\begin{equation}
\label{4}
w_{i} = \alpha \frac{DP_{i} }{DP_{max} } +  \beta  \frac{\overline{SNR_{i} } }{\overline{SNR_{max} }  } + \gamma \frac{storage_{i} }{storage_{max} } 
\end{equation}
where $DP$ is the data processing capacity, $storage$ is the storage capacity, $\alpha $, $\beta $ and $\gamma $ respectively represent the data processing capacity of the RSU, the average SNR, the weight of the storage capacity, and $\alpha +  \beta +  \gamma = 1$.

According to the weight value of the RSU, each RSU will start a timeout clock associated with the weight value. Every time the master node in the intra-group consensus broadcasts a heartbeat packet in the blockchain network in the group, it needs to perform a timeout calculation. The timeout time $T$ obeys the uniform distribution with parameter $\left ( t_{1}, t_{2} + \beta \tau w_{i}  \right )$.
\begin{equation}
\label{5}
T_{i} \sim U\left ( t_{1}, t_{2} + \beta \tau w_{i}  \right ) 
\end{equation}
where, $t_{1} $, $t_{2} $ is the minimum interval of the timeout, $\beta $ and $\tau $ are constants.


When the master node in the intra-group consensus fails, the blockchain system can make the best leader node selection in the group according to the weight value of the node. The leader node elected by the weighted Raft consensus mechanism has better comprehensive capabilities. Whether it is communication ability, data processing capacity, or storage capacity, it belongs to the forefront of the nodes in the intra-group consensus, which helps to improve the consensus efficiency in the group. The replica nodes of the WRBFT algorithm have three different roles: leader, candidate and follower. Table 1 summarizes the symbols and semantic analysis used by the WRBFT algorithm. Algorithm 1 is the intra-group consensus of WRBFT. The intra-group consensus includes three stages: $Leader\textbf{ }Election$, $Block-Proposal$, and $Block-Confirm$.

\begin{algorithm}[H]
\caption{Consensus within the group.}\label{alg:alg1}
\begin{algorithmic}
\STATE 
\STATE $ \triangleright \textbf{ Leader Election}$
\STATE $ \textbf{function }  Heartbeat\left (SNR,DP,storage  \right ) $
\STATE \hspace{0.5cm}$ \overline{SNR}\gets  formula\left ( 2 \right )   $
\STATE \hspace{0.5cm}$ w\gets formula\left ( 3 \right )    $
\STATE \hspace{0.5cm}$ T\gets formula\left ( 4 \right )   $
\STATE \hspace{0.5cm}\textbf{return }  $T$
\STATE $ \textbf{if } $waiting time$ = T \textbf{ then} $
\STATE \hspace{0.5cm}$ $broadcast$ \textbf{ } \left \langle Request-vote, i, v \right \rangle _{\sigma _{i} }  $
\STATE $ \textbf{if } $receive$ \textbf{ } \left \langle  Request-vote,i,v\right \rangle _{\sigma _{i} } \textbf{ }$and$\textbf{ }isVote = $false$ \textbf{ then} $
\STATE \hspace{0.5cm}$ $send$ \textbf{ } \left \langle Reply-vote,i,v \right \rangle _{\sigma _{i} } \textbf{ } $to\textbf{ }candidate$ $
\STATE $ \textbf{while } $receive$ \textbf{ }\left \langle Reply-vote,i,v \right \rangle _{\sigma _{i} } $= $f$ + 1$ \textbf{ do}$
\STATE \hspace{0.5cm}$ $become leader$ $
\STATE $ \triangleright \textbf{ Block-Proposal}$
\STATE $ \textbf{if } $role is leader$\textbf{ then} $
\STATE \hspace{0.5cm}$ $broadcast $ \left \langle \left \langle Block-Proposal,i,v,h \right \rangle _{\sigma _{i} } ,data \right \rangle  $
\STATE $ \textbf{if } $role is follower$\textbf{ then} $
\STATE \hspace{0.5cm}\textbf{if } $ $receive $\left \langle \left \langle Block-Proposal,i,v,h \right \rangle _{\sigma _{i} } ,data \right \rangle \textbf{ then} $ 
\STATE \hspace{0.5cm}\hspace{0.5cm}$ $send $ \left \langle Block-Confirm,i,v,h \right \rangle _{\sigma _{i} } $ to leader$ $
\STATE $ \triangleright \textbf{ Block-Confirm}$
\STATE $ \textbf{if } $role is leader$\textbf{ then} $
\STATE \hspace{0.5cm}\textbf{while } $ $receive $\left \langle Block-Confirm,i,v,h \right \rangle _{\sigma _{i} } $= $f$ + 1$\textbf{ do}$
\STATE \hspace{0.5cm}\hspace{0.5cm}$ $consensus among participating groups$ $
\end{algorithmic}
\label{alg1}
\end{algorithm}

In the $Leader\textbf{ }Election$ phase, the replica nodes in each group start the $Heartbeat(SNR, DP, storage)$ function to calculate the timeout period. When it is found that the leader in the intra-group consensus does not send a heartbeat message within the specified time, the node with a larger weight value has a smaller timeout period, so it is the first to broadcast the $\left \langle Request-vote,i,v \right \rangle _{\sigma _{i} } $ message. It will collect $f$+1 $\left \langle Reply-vote,i,v \right \rangle _{\sigma _{i} } $ messages faster, become the leader, and lead the consensus work of its group.

In the $Block-Propasal$ phase, the leader node in the intra-group consensus packs the data in the storage pool into blocks and broadcasts $\left \langle \left \langle Block-Proposal,i,v,h \right \rangle _{\sigma _{i} },data \right \rangle  $ messages. If the storage pool does not have any vehicle data at this stage, $data = \perp $. After the follower receives the message sent by the leader, it checks the correctness of the message, and after confirming that it is correct, sends an $\left \langle Block-Confirm,i,v,h \right \rangle _{\sigma _{i} } $ message to the leader.

In the Block-Confirm phase, the leader carefully checks the messages sent by the followers. After receiving $f$+1 $\left \langle Block-Confirm,i,v,h \right \rangle _{\sigma _{i} } $ messages, the master node in the intra-group consensus will enter the inter-group consensus.

\section{Inter-group consensus of WRBFT algorithm}
\noindent To make the IoV blockchain system have a certain degree of Byzantine fault tolerance while meeting lower latency, higher throughput, and security, we adopt PBFT algorithm based on BLS aggregated signatures with non-linear coefficients in the inter-group consensus. The PBFT consensus has the problems of high communication complexity and the method of sequentially acting as the master node is vulnerable to targeted attacks by malicious nodes, which limits the application of PBFT in the IoV. We use BLS aggregated signatures to reduce the communication complexity of PBFT to $O\left ( N \right ) $ and use VRF to ensure the randomness of master node selection.

\subsection{BLS Aggregated Signatures with Nonlinear Coefficients}

\noindent The PBFT consensus uses broadcasting to send messages during the prepare and commit phases, resulting in an excessively high information density in the blockchain system. When the number of consensus nodes surges, the information density in the IoV blockchain system increases exponentially. We face an important challenge: how to reduce communication complexity while maintaining the Byzantine elasticity of the PBFT consensus. We apply BLS aggregate signature technology to the prepare and commit stages of PBFT consensus, and the replica node sends the message to the leader node individually. After checking the correctness of the signature, the leader node aggregates 2$f$+1 signatures into an aggregated signature and broadcasts the prepare or commit message containing the aggregated signature to the replica node, effectively reducing the communication complexity of PBFT to $O\left ( N \right ) $.

We introduce non-linear coefficients in the formation of aggregated signatures so that they can resist key attacks such as forged signatures while reducing the complexity of PBFT consensus communication. The BLS aggregate signature method with nonlinear coefficients, as shown in Algorithm 2. The aggregation signature is calculated as follows,
\begin{equation}
\label{6}
\hat{\sigma } =\alpha _{1} \ast \sigma _{1} + \alpha _{2} \ast \sigma _{2} + \cdots + \alpha _{N} \ast \sigma _{N}  
\end{equation}

The calculation method of the aggregated public key is expressed as follows,
\begin{equation}
\label{7}
PK= \alpha _{1} \ast pk_{1}+ \alpha _{2} \ast pk_{2} + \cdots + \alpha _{N} \ast pk_{N}    
\end{equation}

\begin{algorithm}[H]
\caption{Aggregate Signature.}\label{alg:alg2}
\begin{algorithmic}
\STATE 
\STATE $ \textbf{function }  signature\left ( m,sk_{i},H  \right )  $
\STATE \hspace{0.5cm} $\sigma _{i}= sk_{i}\ast H\left ( m \right )   $
\STATE \hspace{0.5cm} \textbf{return }  $\sigma _{i}$
\STATE $ \textbf{function }  factor\left ( pk_{1},pk_{2},\cdots ,pk_{N}    \right )   $
\STATE \hspace{0.5cm} $\alpha _{i}= H\left ( pk_{i} \left |  \right | pk_{1}\left |  \right | pk_{2} \left | \right | \cdots \left |  \right | pk_{N}   \right )   $
\STATE \hspace{0.5cm} \textbf{return }  $\alpha _{i} $
\STATE $ \textbf{function }  BLS\left ( \sigma _{1},\cdots \sigma _{N},pk_{1},\cdots ,pk_{N}     \right )   $
\STATE \hspace{0.5cm} $\hat{\sigma } \gets formula\left ( 5 \right )    $
\STATE \hspace{0.5cm} $PK \gets formula\left ( 6 \right )    $
\STATE \hspace{0.5cm} \textbf{return }  $\hat{\sigma } ,PK$
\STATE $ \textbf{if }  e\left ( \sigma _{i},g_{1}   \right ) = e\left ( pk_{i},H\left ( m \right )   \right ) \textbf{ then}   $
\STATE \hspace{0.5cm} \textbf{return }  true
\STATE $ \textbf{if }  e\left ( \hat{\sigma },g_{1}   \right ) = e\left ( PK,H\left ( m \right )  \right )  \textbf{ then}   $
\STATE \hspace{0.5cm} \textbf{return }  true
\end{algorithmic}
\label{alg2}
\end{algorithm}

\subsection{Inter-group consensus}
\noindent The PBFT consensus method of sequentially selecting the master node is vulnerable to targeted attacks from malicious nodes (replay and desynchronization attacks, etc.) and falls into the process of view switching, resulting in the blockchain system failing to complete block generation within the specified time. The VRF uses cryptography to select the leader node. Before generating a block, other nodes cannot know any information about the leader node in advance, which effectively improves the security of IoV data sharing. We use VRF to ensure the randomness of master node selection. VRF has verifiability, uniqueness, and randomness.

In verifiability, for any key pair $\left ( pk_{i},sk_{i}   \right ) $ and a string $s$ composed of 0/1, if $\left ( \xi _{i},\pi _{i}   \right ) = VRF_{prove}\left ( sk_{i},s  \right )  $, then there is a polynomial $\mu $ which tends to be infinitesimal such that
\begin{equation}
\label{6}
Pr\left [ VRF_{verify}\left ( pk_{i},\xi _{i},\pi _{i},s    \right ) = True  \right ] \\ = 1- \mu \left ( k \right ) 
\end{equation}

In Uniqueness, if the input string $s$ and $sk$ are the same, then the calculated $\xi $ and $\pi $ must be the same, that is, there is no $\left ( \xi _{1},\pi _{1}   \right ) $ and $\left ( \xi _{2},\pi _{2}   \right ) $, so that
\begin{equation}
Pr\left [ \xi _{1} \ne \xi _{2} \mid \begin{array}{c}   VRF_{verify} \left ( pk_{i}, \xi _{1}, \pi _{1}, s \right ) = True \\   VRF_{verify} \left ( pk_{i}, \xi _{2}, \pi _{2}, s \right ) = True\end{array} \right ] \le \mu _{k} 
\end{equation}

In randomness, we don't know $\pi $, there is no difference between $\xi $ and an ordinary random value for malicious nodes.

Each inter-group consensus node in this section calculates the seed value based on the view number and the number of inter-group consensus nodes. We can easily know that the number of consensus nodes between groups is equal to the number of groups of the WRBFT algorithm, and calculate its own $\xi $ and $\pi $, the specific calculation expression is as follows,

\begin{equation}
\label{6}
\left ( \xi _{i}, \pi _{i}  \right ) \gets VRF_{prove}\left ( sk_{i}, seed \right )   
\end{equation}

Then, the inter-group node judges whether it is selected as the master node of this round of consensus according to its random number and the proof of the random number,
\begin{equation}
\label{6}
ElecResult= \left\{\begin{matrix}   True, \left ( \frac{H\left ( \xi _{i}  \right ) }{2^{hashlen} } \le \epsilon  \right )  \\    False, \left ( \frac{H\left ( \xi _{i}  \right ) }{2^{hashlen} } >  \epsilon  \right ) \end{matrix}\right. 
\end{equation}
where, $\epsilon \left ( \epsilon \in \left [ 0, 1 \right ]  \right ) $ is the selection threshold, and $hashlen$ is the hash length.

Due to the nature of VRF, that is, formula (9), only the master node can calculate the proof of the correctness of the random number, and other nodes can only verify the correctness of the proof through formulas (7) (8). Before the master node of inter-group consensus proposes a new block, the replica nodes (including malicious nodes) do not know the identity information of the new round of master nodes, and cannot carry out targeted attacks on the master node.

\begin{algorithm}[H]
\caption{Consensus among groups.}\label{alg:alg3}
\begin{algorithmic}
\STATE 
\STATE $ \triangleright \textbf{ Leader Election}$
\STATE $ \textbf{function }  VRFParm\left ( v, K, sk_{i} \right )   $
\STATE \hspace{0.5cm} $seed\gets v \% K $
\STATE \hspace{0.5cm} $\left ( \xi _{i}, \pi _{i}  \right ) \gets VRF_{prove} \left ( sk_{i}, seed \right ) $
\STATE \hspace{0.5cm} \textbf{return }  $\xi _{i} ,\pi _{i} $
\STATE $ \textbf{if }  $formula(11) = true $ \textbf{ then}   $
\STATE \hspace{0.5cm} become leader
\STATE $ \textbf{function }  verifyLeader\left ( pk, seed, \xi ,\pi ,hashlen \right )   $
\STATE \hspace{0.5cm} $verify\gets VRF\left ( pk, seed, \xi ,\pi  \right ) $
\STATE \hspace{0.5cm} $ElecResult\gets formula\left ( 10 \right )  $
\STATE \hspace{0.5cm} $ \textbf{if } $ $verify$ = true and $ElecResult$ = true$ \textbf{ then} $
\STATE \hspace{0.5cm} \hspace{0.5cm} return true
\STATE \hspace{0.5cm} \textbf{return }  false

\STATE $ \triangleright \textbf{ Pre-prepare}$
\STATE $ \textbf{if }  $role is leader$ \textbf{ then}   $
\STATE \hspace{0.5cm} broadcast $\left \langle \left \langle Pre-prepare, i, v, h \right \rangle _{\sigma _{i} } ,data \right \rangle $

\STATE $ \triangleright \textbf{ Prepare}$
\STATE $ \textbf{if }  $role is leader$ \textbf{ then}   $
\STATE \hspace{0.5cm} \textbf{while }receive $\left \langle Prepare1,i,v,h \right \rangle _{\sigma _{i} } $ = 2$f$+1\textbf{ do}
\STATE \hspace{0.5cm} \hspace{0.5cm} broadcast $\left \langle Prepare,i,v,h \right \rangle_{ \hat{\sigma } } $
\STATE $ \textbf{if }  $role is follower$ \textbf{ then}   $
\STATE \hspace{0.5cm} send $\left \langle Prepare1,i,v,h \right \rangle _{\sigma _{i} } $ to leader

\STATE $ \triangleright \textbf{ Commit}$
\STATE $ \textbf{if }  $role is leader$ \textbf{ then}   $
\STATE \hspace{0.5cm} \textbf{while }receive $\left \langle Commit1,i,v,h \right \rangle _{\sigma _{i} } $ = 2$f$+1\textbf{ do}
\STATE \hspace{0.5cm} \hspace{0.5cm} broadcast $\left \langle Commit,i,v,h \right \rangle_{ \hat{\sigma } } $
\STATE \hspace{0.5cm} \hspace{0.5cm} update blockchain
\STATE $ \textbf{if }  $role is follower$ \textbf{ then}   $
\STATE \hspace{0.5cm} send $\left \langle Commit1,i,v,h \right \rangle _{\sigma _{i} } $ to leader
\STATE \hspace{0.5cm} \textbf{if }receive $\left \langle Commit,i,v,h \right \rangle_{ \hat{\sigma } }  $
\STATE \hspace{0.5cm} \hspace{0.5cm} update blockchain
\end{algorithmic}
\label{alg3}
\end{algorithm}

Algorithm 3 is the inter-group consensus of WRBFT. The inter-group consensus includes four stages: $Leader\textbf{ }Election$, $Pre-prepare$, $Prepare$, and $Commit$. We apply the BLS aggregate signature to the $Prepare$ and $Commit$ phases, and these two phases are subdivided into two sub-phases to complete the work of consensus within the group.

\begin{figure}[!t]
\centering
\includegraphics[width=3.5in]{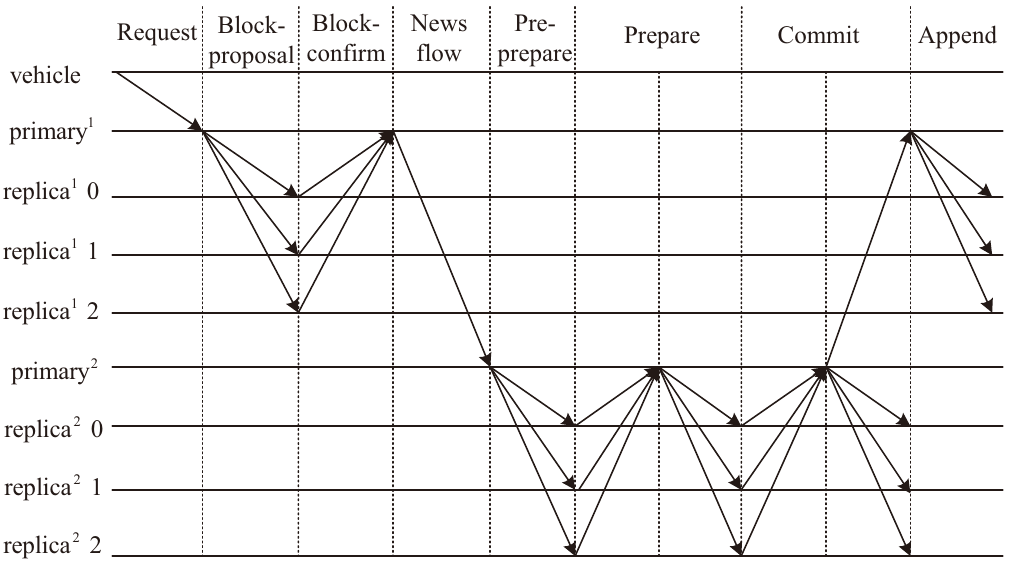}
\caption{The implementation flow chart of the WRBFT algorithm (the superscripts of primary and replica represent the layer number, the first layer represents the intra-group consensus, and the second layer represents the inter-group consensus).}
\label{fig_1}
\end{figure}

In the $Leader Election$ phase, the master nodes of each group in the intra-group consensus enter the inter-group consensus, and the locally optimal master nodes participate in the global consensus. The delay of WRBFT will have a lower delay and higher throughput. The inter-group nodes first execute the $VRFParm\left ( v,R,sk_{i}  \right ) $ function during leader election to obtain the random number $\xi _{i} $ and the random number proof $\pi _{i} $, and then use formula (10) to calculate whether they meet the leader threshold requirements. The inter-group consensus node that meets the requirements will become the master node to lead the current round of inter-group consensus work after being approved by 2$f$+1 followers.

In the $Pre-prepare$ phase, the leader of inter-group consensus sorts the received vehicle data, and broadcasts $\left \langle \left \langle Pre-prepare,i,v,h \right \rangle _{\sigma _{i} }, data \right \rangle $ when the maximum waiting time is reached or the maximum amount of data that the block can carry is collected.

In the $Prepare$ phase, the $Prepare$ phase of the inter-group consensus is divided into two sub-phases, which avoids the problem of excessive message density caused by the broadcast of each node in the $Prepare$ phase of PBFT. In the first sub-phase, after receiving the $Pre-prepare$ message from the leader, the follower sends $\left \langle Prepare1,i,v,h \right \rangle _{\sigma _{i} } $ to the leader alone. The leader checks the signature of the follower message, and after receiving 2$f$+1 accurate $\left \langle Prepare1,i,v,h \right \rangle _{\sigma _{i} } $ messages, aggregates 2$f$+1 signed messages into one signature. In the second subphase, the leader broadcasts $\left \langle Prepare,i,v,h \right \rangle _{\hat{\sigma } } $. By using aggregated signatures in the $Prepare$ phase, the communication complexity of this phase is reduced to $O\left ( N \right ) $.

In the $Commit$ phase, the $Commit$ phase of the inter-group consensus is similar to the $Prepare$ phase and is also divided into two sub-phases. In the first sub-phase, the follower receives the $\left \langle Prepare,i,v,h \right \rangle _{\hat{\sigma } } $, confirms that the aggregate signature is correct, and then sends the $\left \langle Commit1,i,v,h \right \rangle _{\sigma _{i} } $ to the leader. The leader integrates 2$f$+1 correct signatures into an aggregate signature and broadcasts $\left \langle Commit,i,v,h \right \rangle _{\hat{\sigma } } $ in the second phase. After that, all nodes of the WRBFT algorithm append new blocks to the blockchain. The implementation flowchart of WRBFT algorithm is shown in Fig.3.

\section{EVALUATION}
\subsection{Configuration}

\begin{figure}[!t]
\centering
\includegraphics[width=3in]{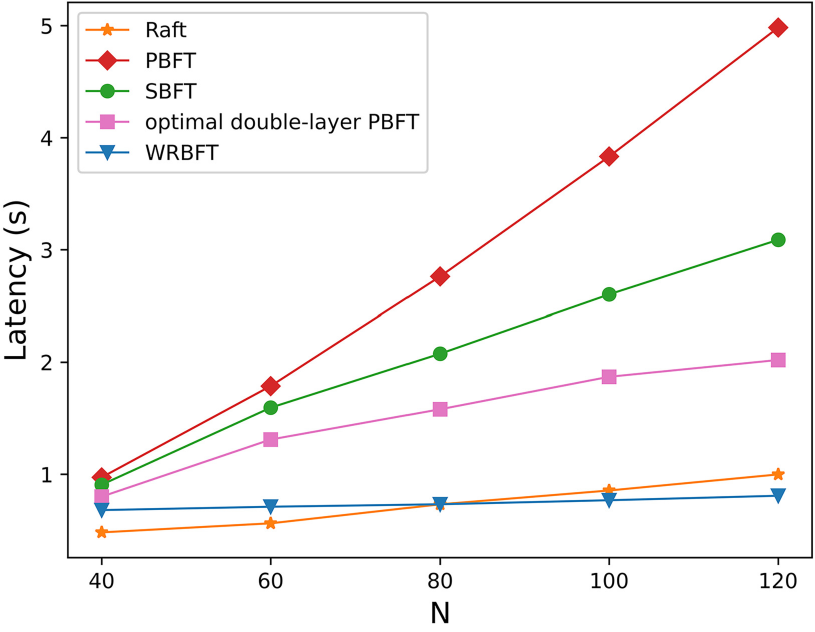}
\caption{Consensus latency of five consensus algorithms.}
\label{fig_1}
\end{figure}


\begin{figure}[!t]
\centering
\includegraphics[width=3in]{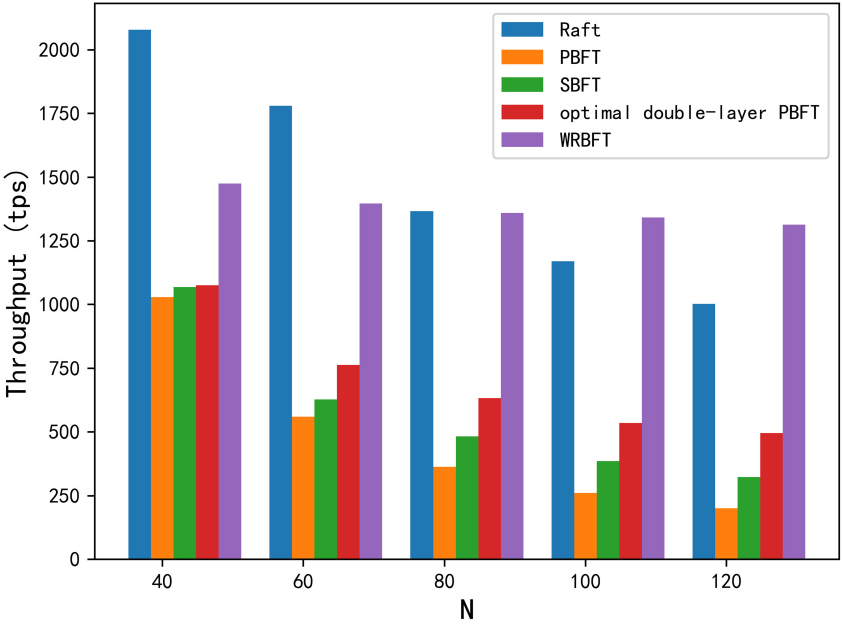}
\caption{Throughput of five consensus algorithms.}
\label{fig_1}
\end{figure}


\begin{figure}[!t]
\centering
\includegraphics[width=3in]{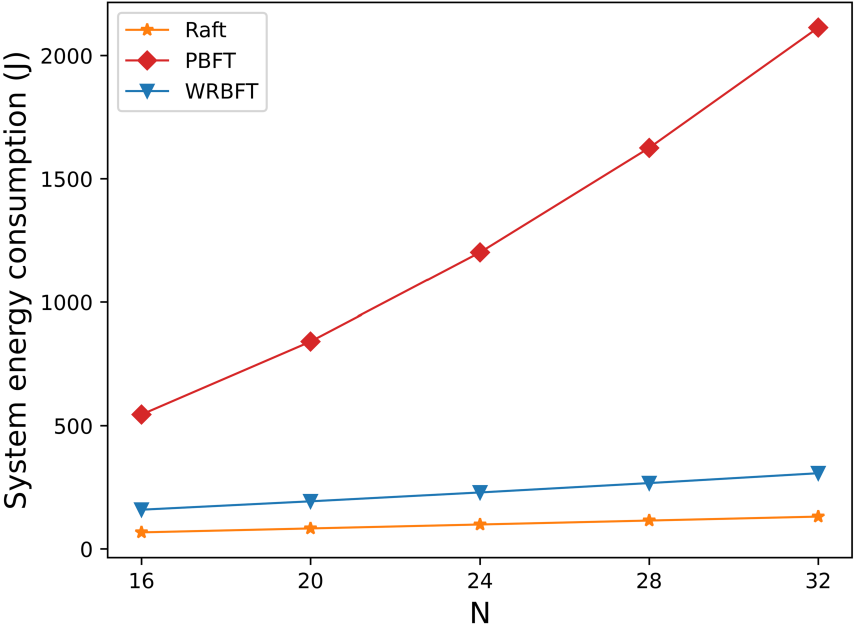}
\caption{System energy consumption of WRBFT algorithm.}
\label{fig_1}
\end{figure}

\begin{figure}[!t]
\centering
\includegraphics[width=3in]{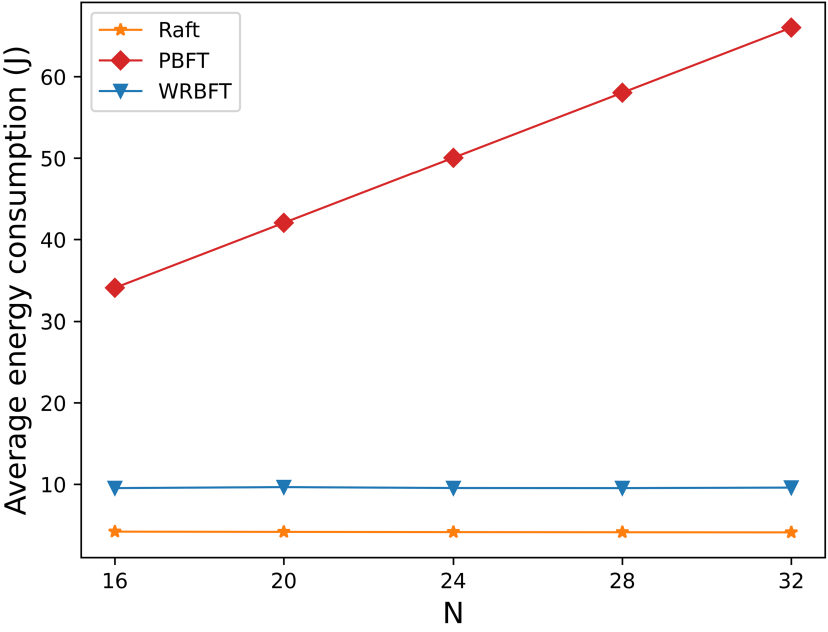}
\caption{Average energy consumption of WRBFT algorithm.}
\label{fig_1}
\end{figure}

\begin{figure}[!t]
\centering
\includegraphics[width=3in]{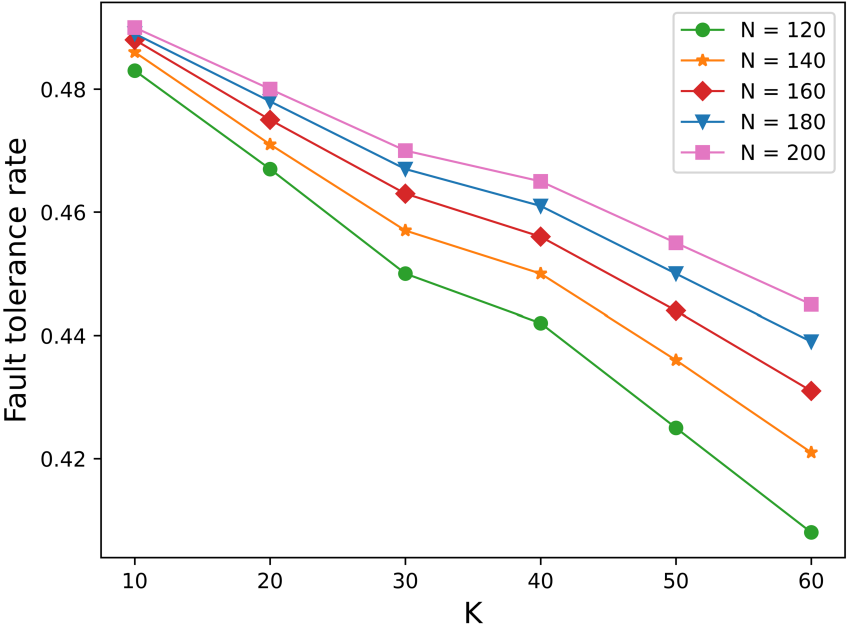}
\caption{Fault tolerance rate and group number of WRBFT algorithm.}
\label{fig_1}
\end{figure}

\begin{figure}[!t]
\centering
\includegraphics[width=3in]{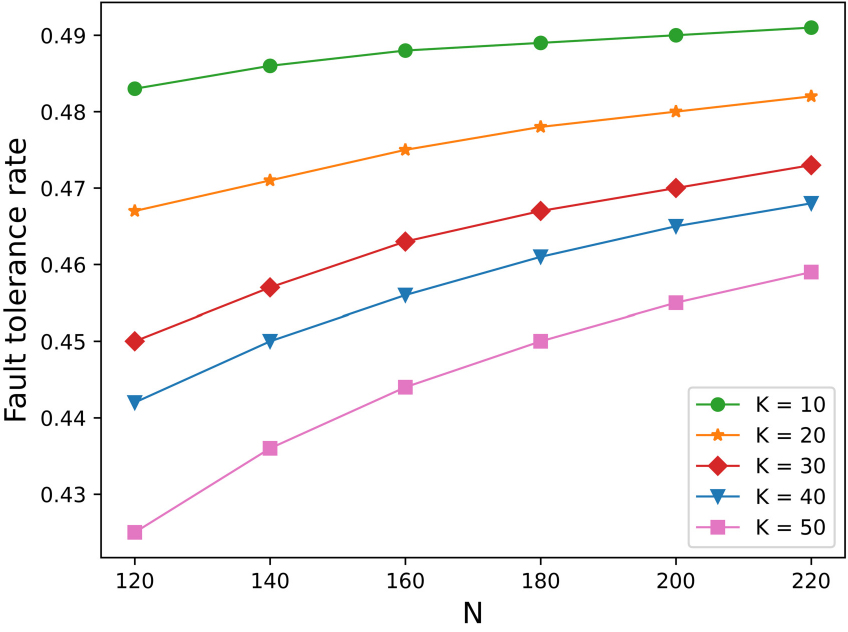}
\caption{Fault tolerance rate and total number of nodes of WRBFT algorithm.}
\label{fig_1}
\end{figure}

\noindent We simulated the WRBFT, Raft, PBFT, SBFT\cite{ref43}, and optimal double-layer PBFT \cite{ref13} consensus algorithms using Python code, and conducted experiments on a CPU (i5-10210U) with 8 cores (1.60GHz) and 16GB RAM. The WRBFT algorithm uses the Pypbc library to implement the process of ordinary signatures and BLS aggregate signatures. The consensus node generates a key pair through the KeyGen function under the Pypbc library, creates a genesis block, establishes a node connection, and exchanges the node IP address list to complete the guidance of the blockchain consensus. We will evaluate the performance of the WRBFT algorithm through four standard indicators of consensus latency, throughput, energy consumption, and fault tolerance rate.

\subsection{Evaluation Results}

\noindent \textbf{Consensus Latency: }We define consensus latency as the time difference from the block proposal to the final confirmation of the block on the chain. We conduct 50 consecutive experiments, and the following indicators run the same number of experiments, and take the average consensus delay as the system consensus delay. For simplicity, we default the number of groups to 4 for WRBFT and optimal double-layer PBFT. Fig. 4 shows the consensus delay comparison between WRBFT and Raft, PBFT, SBFT, and optimal double-layer PBFT. As the number of nodes increases, the delay of all consensus algorithms gradually increases, and due to the high complexity of PBFT communication, the consensus delay is the highest.

When the number of nodes is less than 80, the consensus delay of Raft is the lowest, but when the total number of nodes exceeds 80, the consensus delay of WRBFT reaches the lowest. Because when there are fewer nodes, the inter-group consensus delay of the WRBFT algorithm accounts for a larger proportion of the total delay, and the delay is larger than that of the Raft algorithm. However, as the number of nodes increases and the number of groups remains unchanged, the proportion of inter-group delay decreases.

\begin{figure}[!t]
\centering
\includegraphics[width=3in]{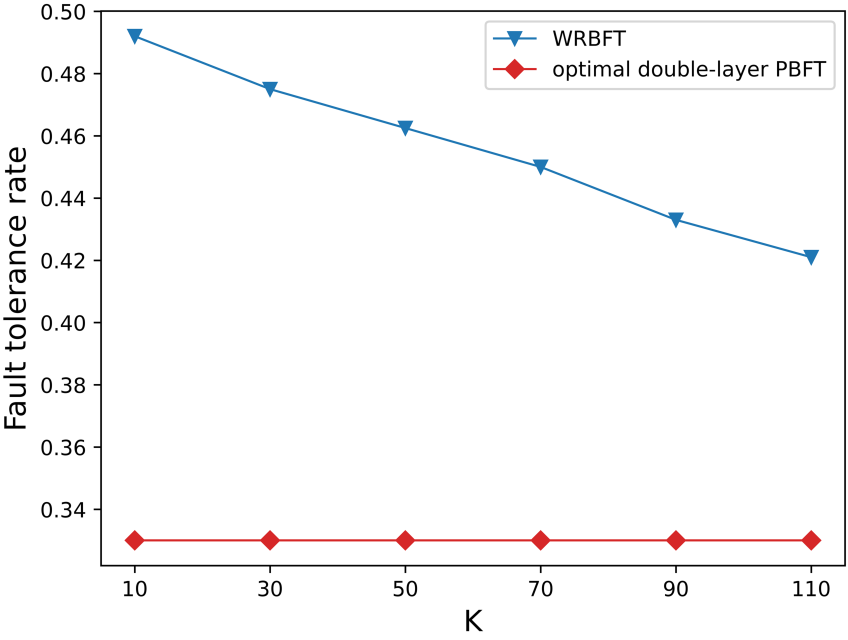}
\caption{Fault tolerance rate and number of groups.}
\label{fig_1}
\end{figure}


\begin{figure}[!t]
\centering
\includegraphics[width=3in]{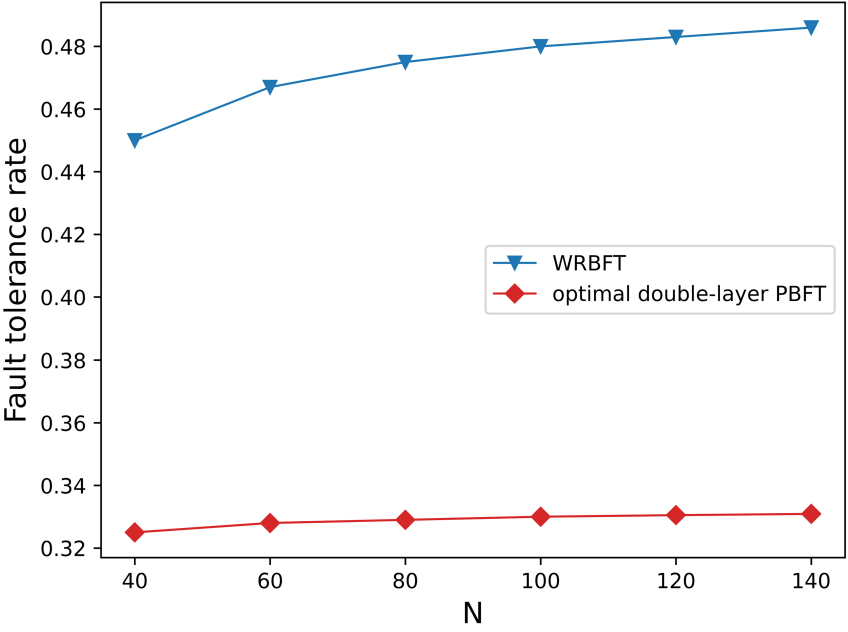}
\caption{Fault tolerance rate and total number of nodes.}
\label{fig_1}
\end{figure}

\textbf{Throughput: }We compared the throughput of Raft, PBFT, SBFT, optimal double layer PBFT, and WRBFT algorithms, and set the block data volume to 2000, as shown in Fig. 5. The throughput of WRBFT and Raft is high. Since the throughput of the WRBFT algorithm decreases slowly, the throughput advantage becomes more obvious when there are more nodes.

\textbf{Energy consumption: }The WRBFT algorithm we proposed is mainly used in the IoV, and energy consumption will be an important indicator to measure the performance of the algorithm. In this section, the system energy consumption and the average energy consumption of nodes are used to evaluate the blockchain consensus algorithm. The energy consumption of the blockchain system mainly comes from the amount of message forwarding and hash times of the consensus nodes. In Fig. 6, since the communication complexity of the PBFT algorithm is polynomial level, the number of messages of consensus nodes is large and accounts for a large proportion of system energy consumption, and its system energy consumption is the largest. The system energy consumption of the Raft and WRBFT algorithms increases slowly as the number of nodes increases. In the WRBFT algorithm, VRF anonymously selects the master node mechanism involves a large number of hash calculations, so the system energy consumption is larger than that of Raft, but it is similar to the system energy consumption of the Raft algorithm. Fig. 7 shows the changing trend of the average energy consumption of nodes with different algorithms. As the number of consensus nodes increases, the message volume of the PBFT algorithm will overwhelm the entire blockchain network.

\begin{figure}[!t]
\centering
\includegraphics[width=3in]{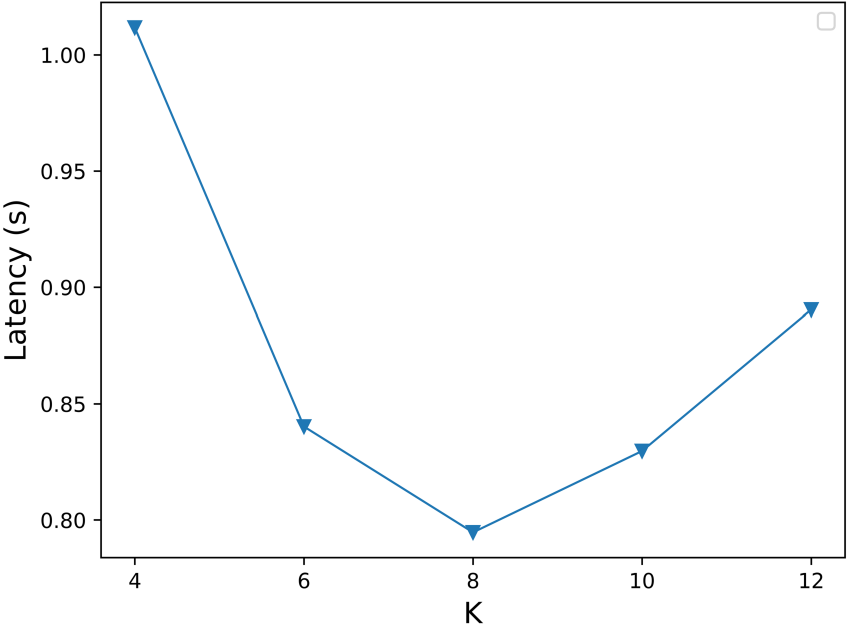}
\caption{WRBFT algorithm consensus latency and number of groups.}
\label{fig_1}
\end{figure}

\begin{figure}[!t]
\centering
\includegraphics[width=3in]{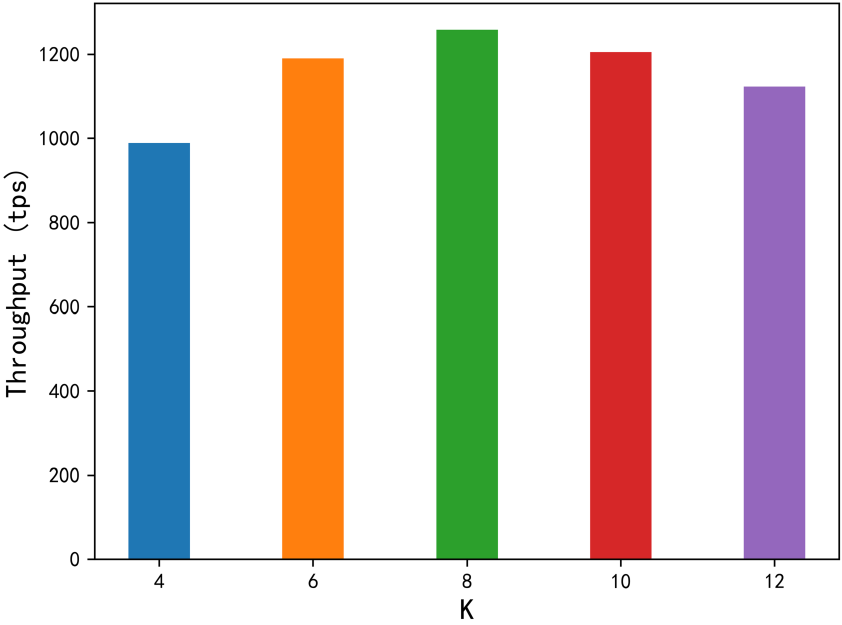}
\caption{WRBFT algorithm throughput and number of groups.}
\label{fig_1}
\end{figure}

\textbf{Fault tolerance rate: }Then we evaluated the performance of WRBFT in the fault tolerance rate in Fig. 8-11, and we proposed that the number of fault-tolerant nodes of the WRBFT algorithm is 
\begin{equation}
\label{7}
F\le - \frac{K}{6} +  \frac{N}{2} - \frac{1}{3}    
\end{equation}
As shown in Fig. 8, we investigate the effect of K on the error tolerance rate of WRBFT when other parameters are fixed. When the total number of nodes is fixed, as K increases, the fault tolerance rate decreases. When the number of groups is 1, the fault tolerance rate of WRBFT is the same as that of Raft, with a fault tolerance rate of 50\%; when the number of groups is equal to the number of nodes, the fault tolerance rate of WRBFT is the same as that of PBFT, with a fault tolerance rate of 33\%. Fig. 9 shows the relationship between the WRBFT fault tolerance rate and the total number of nodes, and the fault tolerance rate is positively correlated with the total number of nodes. It can be seen from Fig. 10 and Fig. 11 that the error tolerance rate of the WRBFT algorithm is greater than that of the optimal double-layer PBFT because the optimal double-layer PBFT group adopts the PBFT consensus between groups.

We also tested the consensus delay and throughput of the WRBFT algorithm under different grouping strategies. The number of nodes tested in Fig. 12 and Fig. 13 was 240, which were divided into 4, 6, 8, 10, and 12 groups to test the consensus delay. According to the changing trend, the analysis shows that when the number of groups is 8 to 10, the consensus delay is the smallest and the throughput reaches the peak.

In short, compared with the Raft consensus, the WRBFT algorithm increases Byzantine elasticity while the consensus latency is almost close, and can tolerate up to 1/3 of malicious nodes. Due to the use of BLS aggregated signatures, the WRBFT algorithm has lower communication complexity than PBFT and optimal double-layer PBFT consensus. At the same time, compared with the SBFT consensus and Hotstuff consensus, the WRBFT algorithm has low latency and high scalability. Therefore, in IoV data sharing, the WRBFT algorithm not only has low latency, high throughput, and high security but also has Byzantine elasticity and high scalability. Table 2 shows the comparison between WRBFT and the most advanced consensus. We can choose different consensus mechanisms according to different needs.

\begin{table*}[!t]
\caption{Performance Comparisons of the Proposed and State-of-the-Art Consensuses \label{tab:table1}}
\centering
\begin{tabular}{ccccc}
\hline \textbf{ } & \textbf{Byzantine fault tolerance} & \textbf{Latency} & \textbf{Communication complexity} & \textbf{Scalability} \\
\hline 
PBFT \cite{ref11} & Yes & High & $O\left ( N^{2}  \right ) $  & Low\\
Raft \cite{ref12} & No & Low & $O\left ( N \right ) $ & High \\
SBFT \cite{ref43} & Yes & Medium & $O\left ( N \right ) $ & Medium \\
Hotstuff \cite{ref44} & Yes & Medium & $O\left ( N \right ) $ & High \\
Optimal double-layer PBFT \cite{ref13}  & Yes & Medium & $1.9N^{\frac{4}{3} } $ & Medium \\
WRBFT (proposed) & Yes & Low & $O\left ( K \right ) +  O\left ( \frac{N}{K}  \right ) $ & High\\
\hline
\end{tabular}
\end{table*}

\section{Conclusion}
\noindent This paper proposes an Internet of Vehicles data-sharing algorithm based on the master-slave consortium chain - WRBFT, which has the characteristics of low latency, high throughput, and high security, and is suitable for safe and efficient data sharing in the Internet of Vehicles. The WRBFT algorithm selects the optimal leader node in the group by comprehensively evaluating the average SNR, data processing capacity, and storage capacity of the nodes in the group, which improves the efficiency of the blockchain system. The WRBFT uses technologies such as BLS aggregation signature and VRF with nonlinear coefficients between groups, which can effectively resist malicious attacks such as key attacks, replay attacks, and desynchronization attacks of malicious nodes while reducing the complexity of PBFT consensus communication. A large number of experimental results show that the WRBFT algorithm effectively reduces the delay and energy consumption, and improves the throughput. In future research, dynamic grouping can be used to balance the delay and throughput within the group to further improve consensus efficiency.


\begin{IEEEbiography}[{\includegraphics[width=1in,height=1.25in,clip,keepaspectratio]{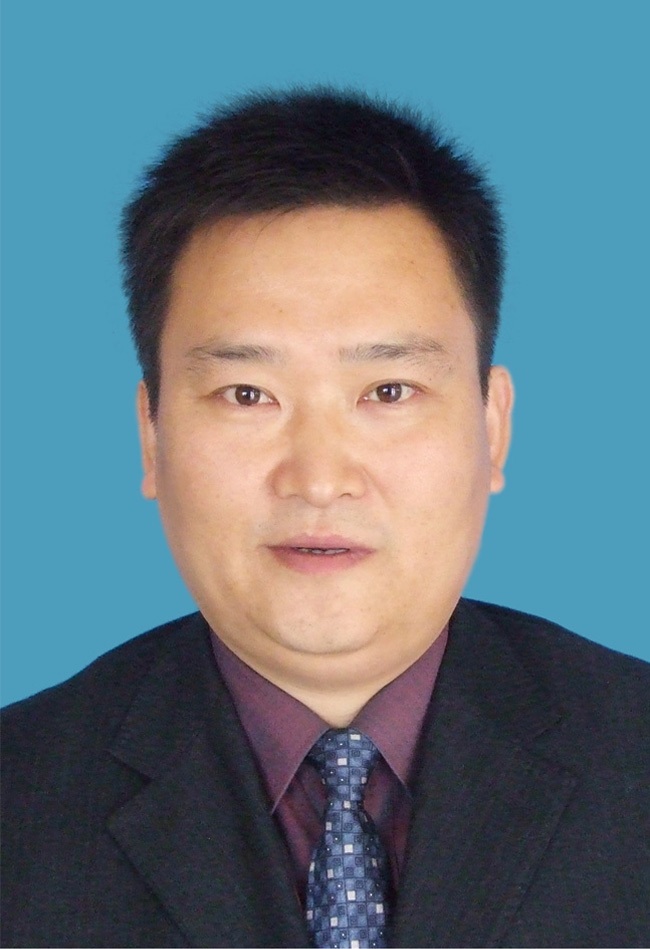}}]{Feng Zhao}
 received the Ph.D. degree in communication and information systems from Shandong University, China, in 2007. He is currently a Professor with the Guangxi Engineering Research Center of Industrial Internet Security and Blockchain, Guilin University of Electronic Technology, Guilin, China. His research interests include cognitive radio networks, MIMO technologies, cooperative communications, and information security.
\end{IEEEbiography}

\begin{IEEEbiography}[{\includegraphics[width=1in,height=1.25in,clip,keepaspectratio]{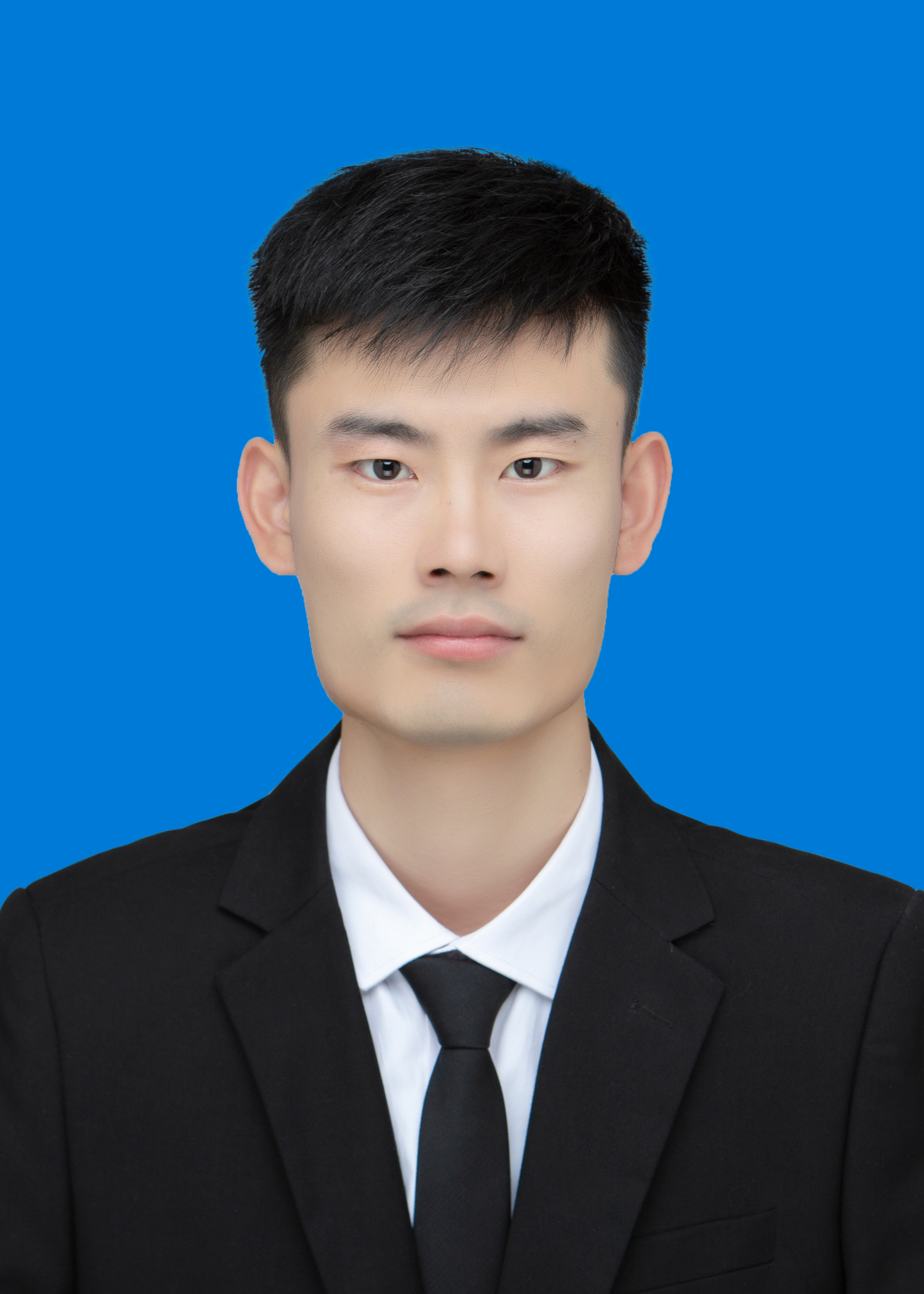}}]{Benchang Yang}
 received the Bachelor's degress from Shandong Agriculture And Engineering University, Jinan, China, in 2020. He is working toward a Master's degree with the Guangxi Engineering Research Center of Industrial Internet Security and Blockchain, Guilin University of Electronic Technology, Guilin, China. His research interests include blockchain and federated learning.
\end{IEEEbiography}

\begin{IEEEbiography}[{\includegraphics[width=1in,height=1.25in,clip,keepaspectratio]{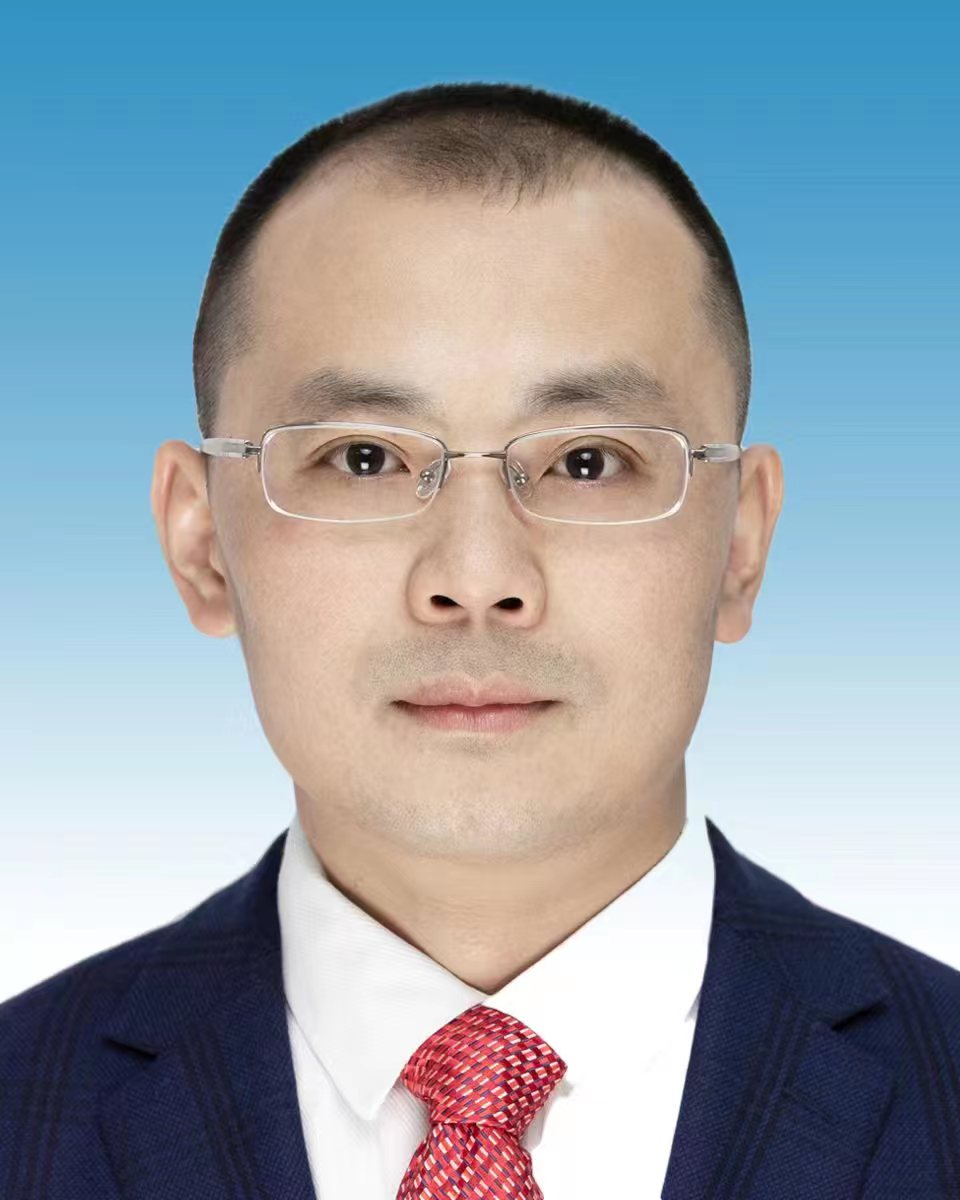}}]{Chunhai Li}
 received the Ph.D. degree from Guilin University of Electronic Technology, Guilin, China, in 2020. He is a Professor with the Guangxi Engineering Research Center of Industrial Internet Security and Blockchain, Guilin University of Electronic Technology. His current research interests include Internet of things security, network security and blockchain.
\end{IEEEbiography}

\begin{IEEEbiography}[{\includegraphics[width=1in,height=1.25in,clip,keepaspectratio]{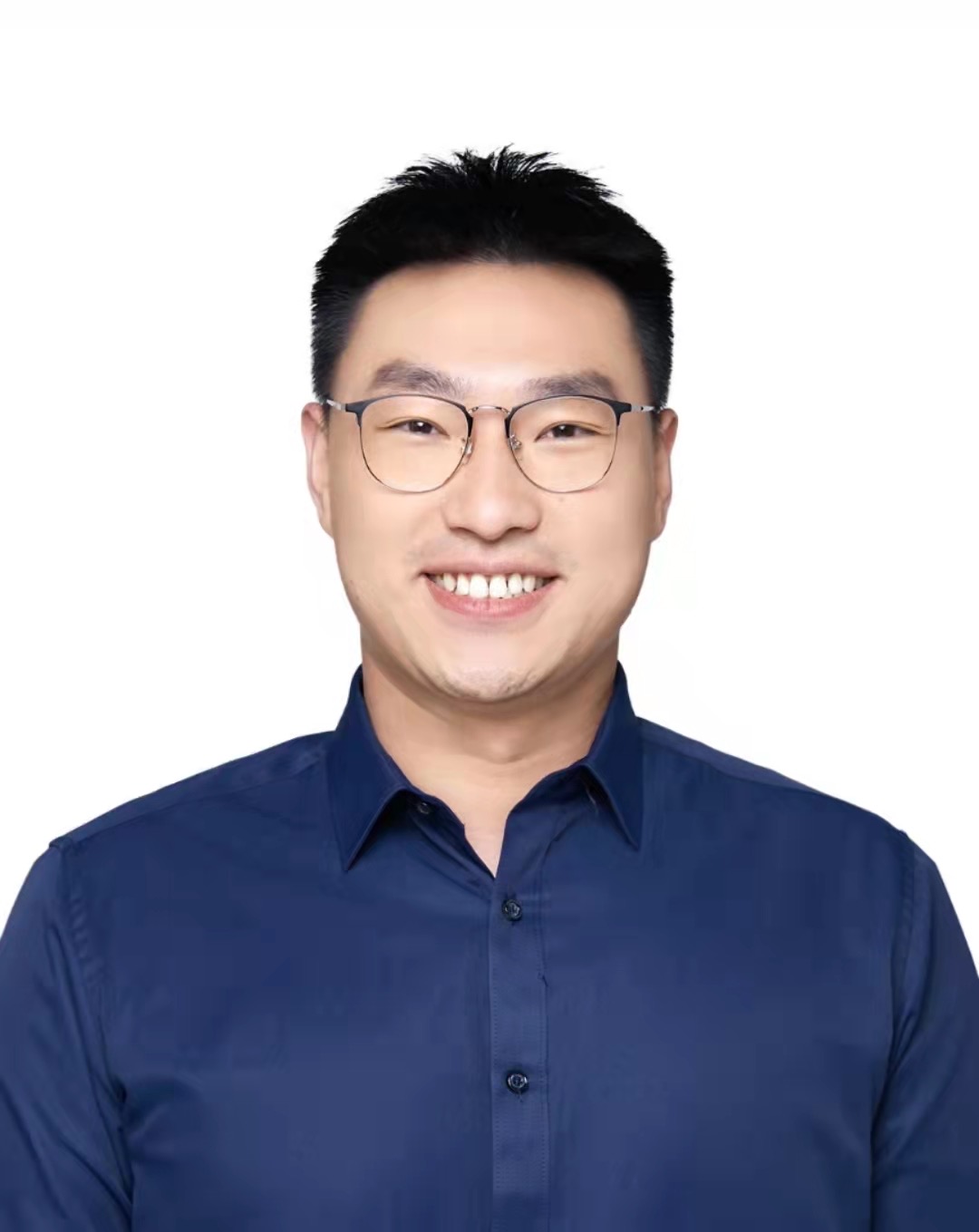}}]{Chuan Zhang}
 received his Ph.D. degree in computer science from Beijing Institute of Technology, Beijing, China, in 2021. From Sept. 2019 to Sept. 2020, he worked as a visiting Ph.D. student with the BBCR Group, Department of Electrical and Computer Engineering, University of Waterloo, Canada. He is currently an assistant professor at the School of Cyberspace Science and Technology, Beijing Institute of Technology. His research interests include secure data services in cloud computing, applied cryptography, machine learning, and blockchain.
\end{IEEEbiography}

\begin{IEEEbiography}[{\includegraphics[width=1in,height=1.25in,clip,keepaspectratio]{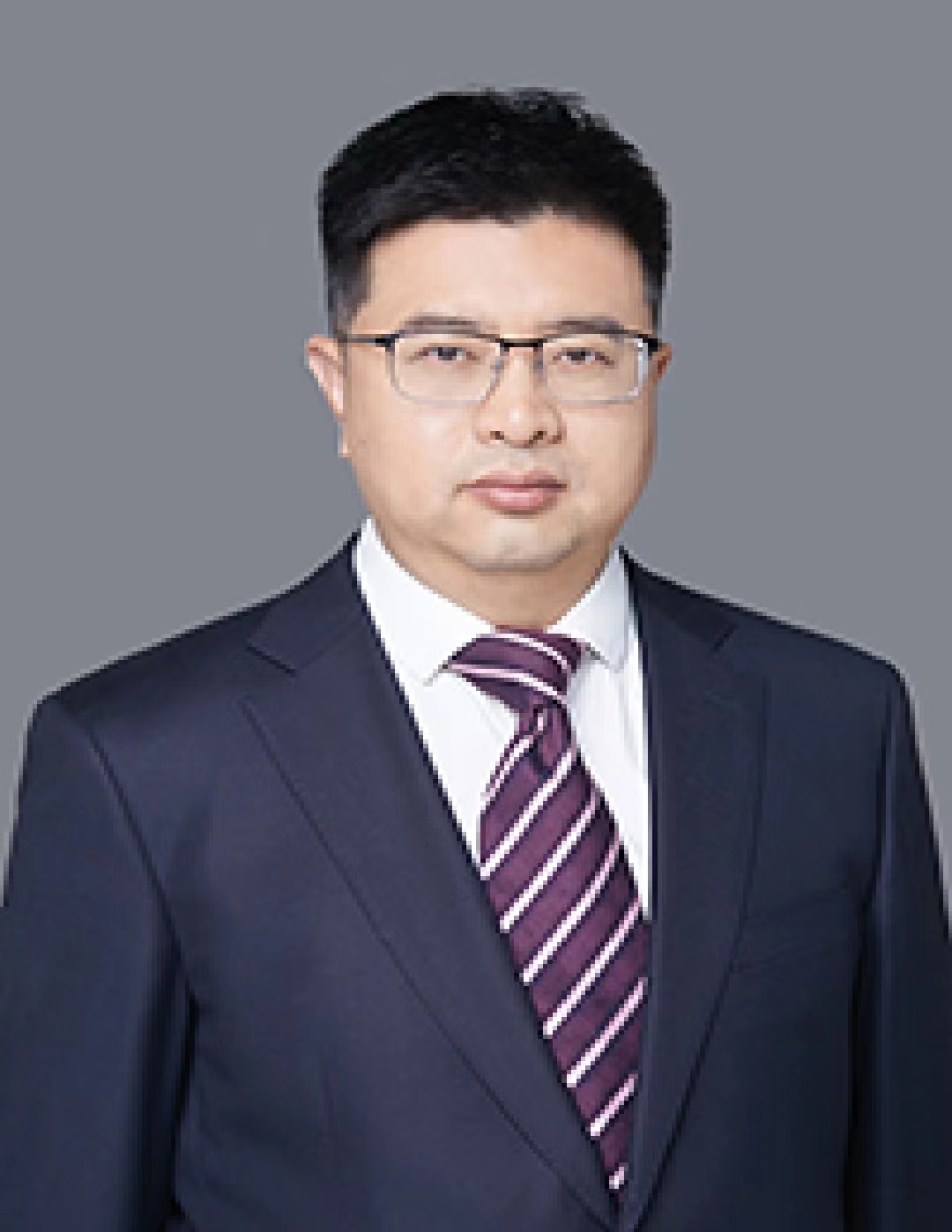}}]{Liehuang Zhu}
 received his Ph.D. degree in computer science from Beijing Institute of Technology, Beijing, China, in 2004. He is currently a professor at the School of Cyberspace Science and Technology, Beijing Institute of Technology. His research interests include security protocol  analysis and design, group key exchange protocols, wireless sensor networks, cloud computing, and blockchain applications.
\end{IEEEbiography}

\begin{IEEEbiography}[{\includegraphics[width=1in,height=1.25in,clip,keepaspectratio]{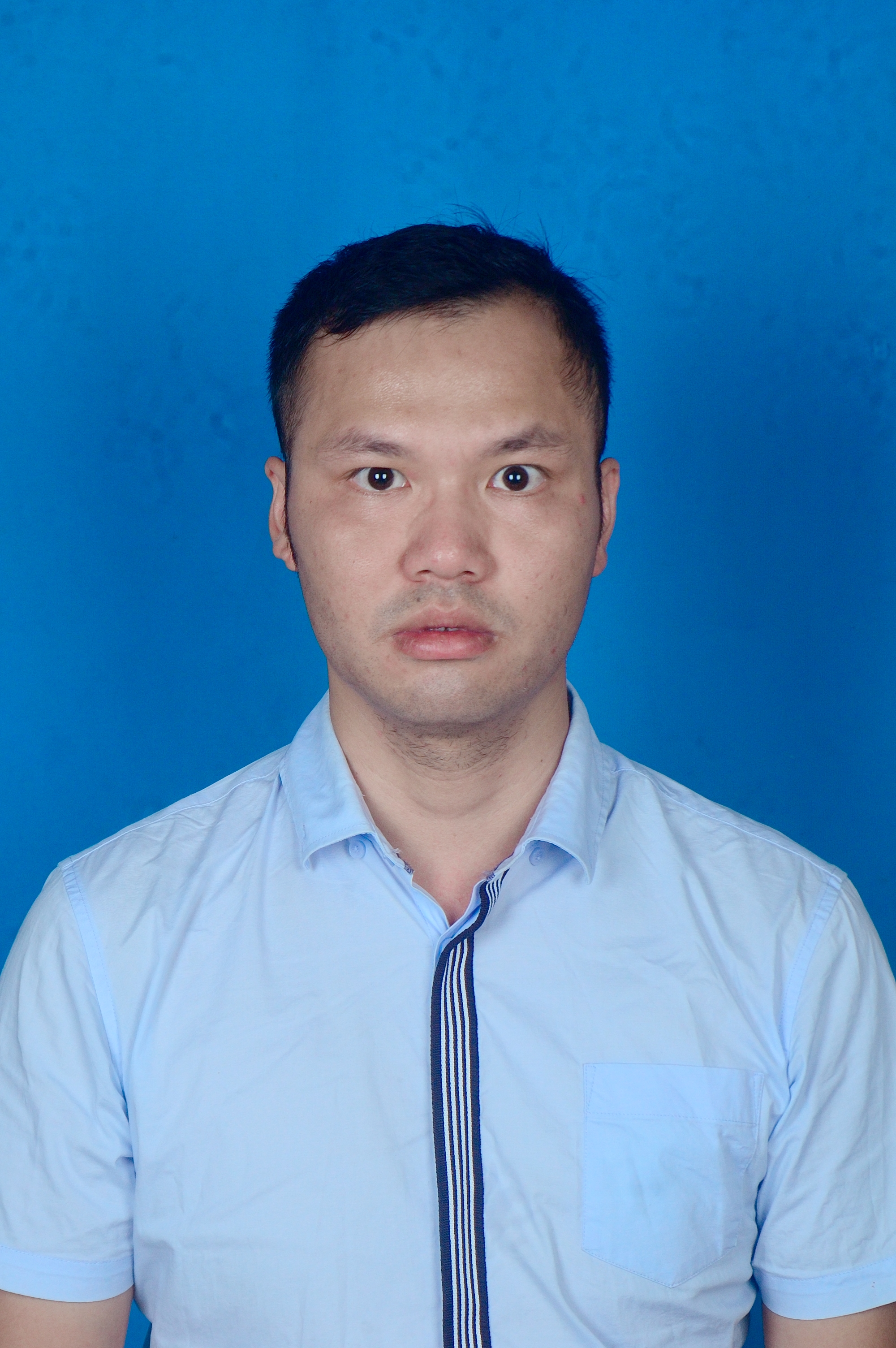}}]{Guoling Liang}
 received his Bachelor's and Master's degrees from Guilin University of Electronic Technology in China in 2012 and 2015, respectively, where he is currently pursuing the doctoral degree. He is currently a teacher in the School of Physics and Telecommunication Engineering at Yulin Normal University. His current research interests include wireless blockchain and edge computing.
\end{IEEEbiography}

\vfill

\end{document}